\documentclass{camera}
\usepackage{graphicx}  
\usepackage{deluxetable}
\begin{document}
%
%
\title{GRB~090323 and GRB~090328: \\ two long high--energy GRBs detected with Fermi
}

%
\author{E.~Bissaldi \\ {\small{on behalf of the Fermi/GBM Team}}}

%
\organization{Max--Planck--Institut f\"ur extraterrestrische Physik, \\ 
           Giessenbachstrasse 1, 85748 Garching, Germany}

\maketitle

\begin{abstract}
We present the analysis of two long Gamma--Ray Bursts, GRB 090323 and GRB
090328, which triggered the Fermi Gamma--Ray Burst Monitor (GBM) and generated 
an Autonomous Repoint Request to the Fermi Large Area Telescope. 
The GBM light curves show multi--peaked structures for both events. 
Here, we present time--integrated and time--resolved burst spectra fitted with 
different models by the GBM detectors.
\end{abstract}

\section{Introduction}
The Fermi Gamma--ray Space Telescope
is an international and multi--agency space observatory,
whose payload comprises two science
instruments, the Large Area Telescope (LAT, 20 Mev--300 GeV)
\cite{ATW09} and the 
GBM \cite{MEE09}.
The primary role of GBM is to augment the science
return from Fermi in the study of Gamma-Ray Bursts (GRBs) by discovering 
transient events within a larger field of view (FoV) and 
performing time-resolved spectroscopy of the measured 
burst emission.
The GBM detectors comprise 12 thallium activated
sodium iodide (NaI(Tl)) scintillation detectors
(8--1000 keV) and two bismuth
germanate (BGO) scintillation detectors
(0.2--40 MeV).
The individual NaI detectors are mounted around the 
spacecraft and their axes are oriented such that the
positions of GRBs can be derived from the measured relative
counting rates. 
The two BGO detectors 
are mounted on opposite sides of the Fermi spacecraft,
covering a net $\sim$8 sr FoV, and provide the overlap in 
energy with the LAT instrument.
\section{Observations of GRB~090323 and GRB~090328}
As of December 11, 2009, the LAT detected 14 GRBs at energies
above 100 MeV. In the following sections, we report the observations 
and analysis of gamma--ray emission from
two GRBs detected within a week of each other in late
March 2009, namely GRB~090323 and GRB~090328.
These bursts have several interesting features.
They were bright enough to trigger an autonomous repointing
of the spacecraft, thus allowing observations by the LAT
for five hours (subject to Earth avoidance).
Moreover, the bursts were detected by the 
LAT Automated Science Processing (ASP) 
by using 6 hours of data.
The improved locations obtained by the LAT instruments
made it possible to follow--up the bursts in the X--ray and in the optical,
thus providing spectroscopic redshifts of $z=3.6$ \cite{CHO09},
and $z=0.736$ \cite{CEN09}, respectively. 
Moreover, there is evidence of emission detected in the LAT up to late times,
which will be extensively discussed in \cite{ABD10}.
Here, we will mainly focus on the GBM analysis of these events,
including temporal and spectral properties, both
time--integrated and time--resolved.
%
%
\begin{figure}[t!]
\centering
\begin{tabular}{c}
\includegraphics[width=0.6\textwidth,bb=30 2 590 554,clip]{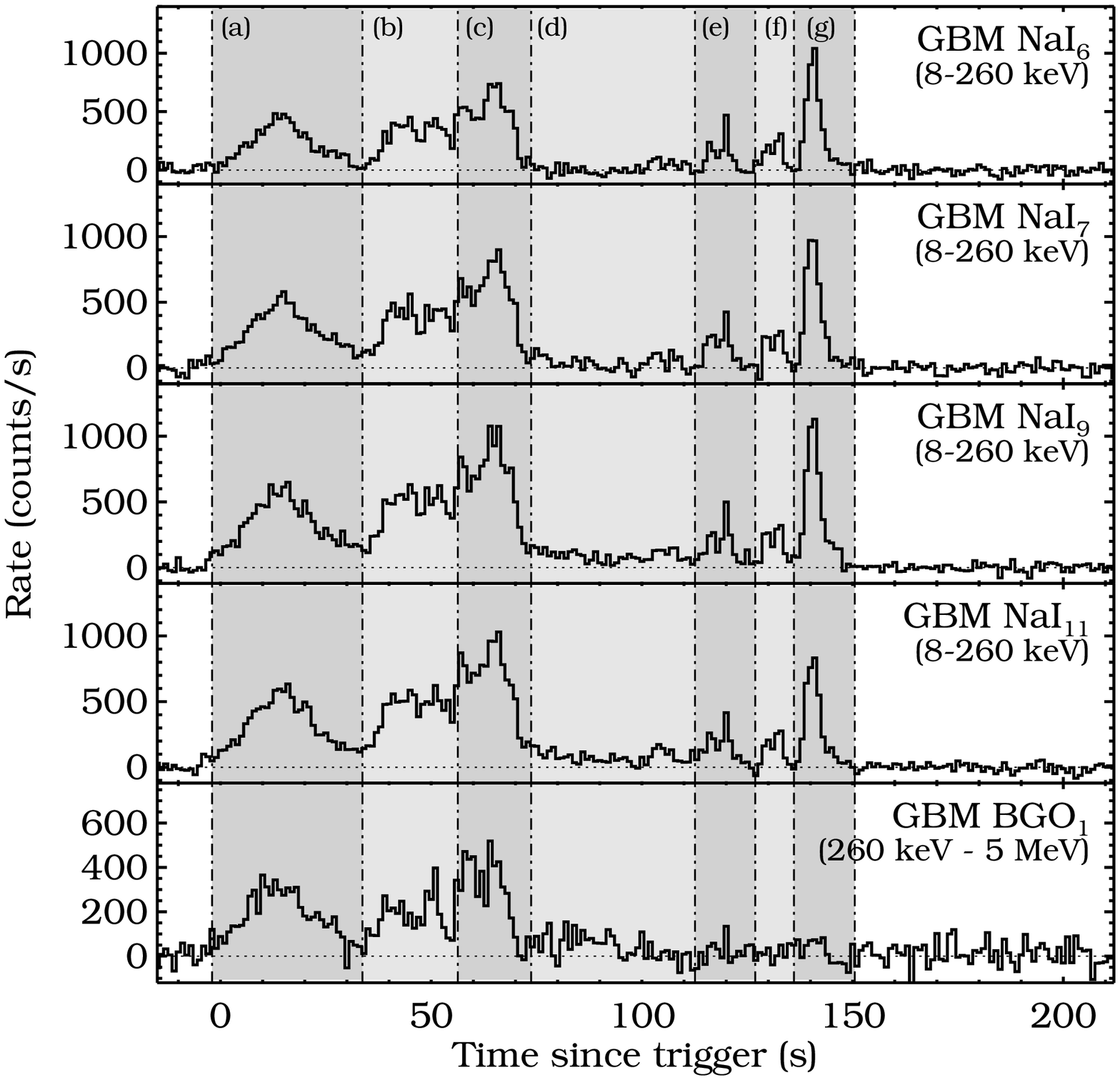} \\
\includegraphics[width=0.6\textwidth,bb=30 2 590 364,clip]{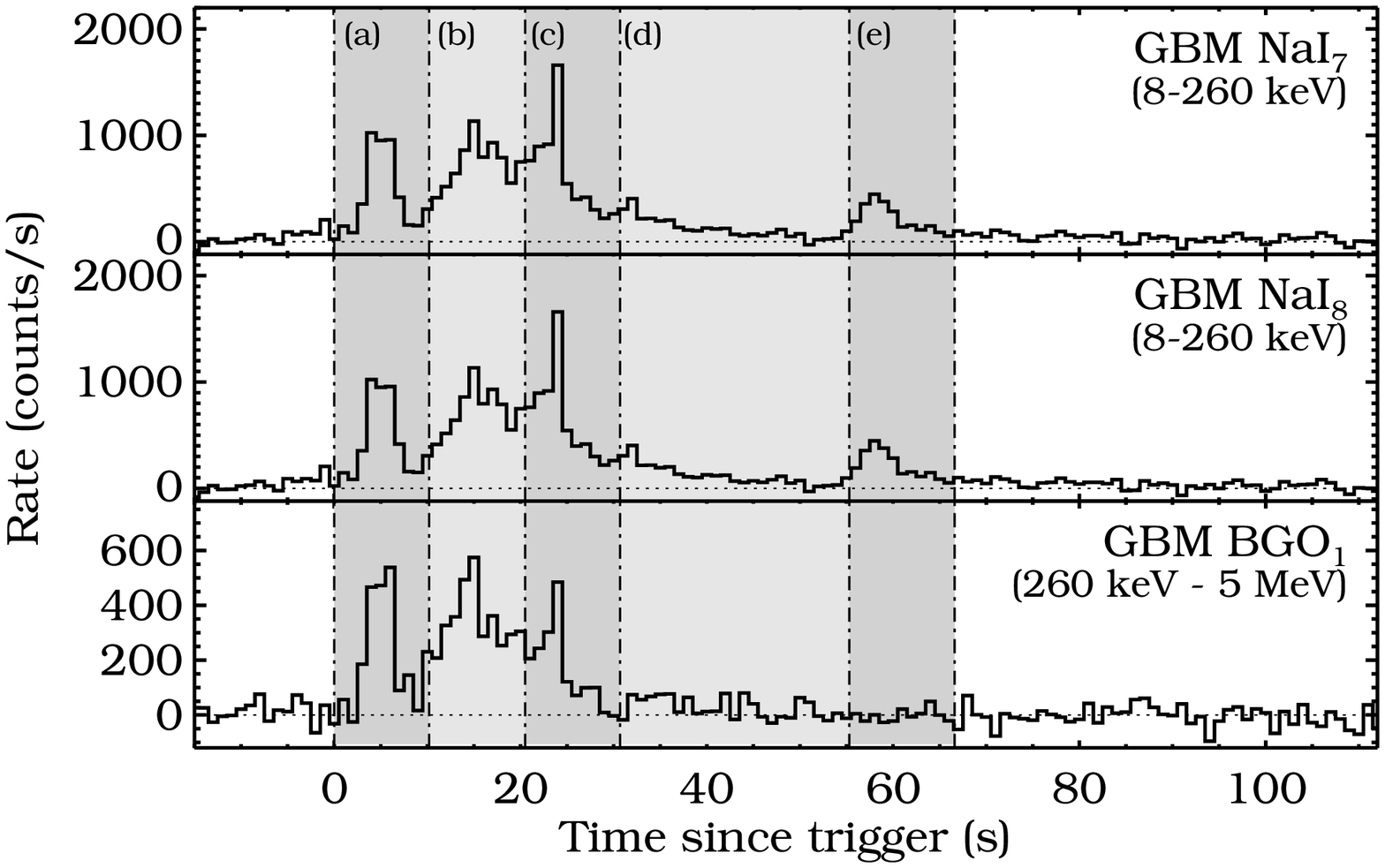} 
\end{tabular}
\caption{Light curves of GRB~090323 ({\it top})
and GRB~090328 ({\it bottom}) observed by the GBM detectors, 
from lowest to highest energies.
The {\it top four (two) panels} show the background--subtracted
count rates, in the 8--260 keV energy band, 
of the four (two) most illuminated NaI detectors.
For both bursts, the {\it bottom panel} is the corresponding plot for
BGO detector 1, between 260 keV and 5 MeV.
The vertical dash--dotted lines indicate the time
intervals used in the time--resolved
spectral analysis.
In all cases, the bin width is 1~s.}
\label{Fig_LC_090323}
\end{figure}
%
%
\section{GBM light curves}
On 2009 March 23 at 00:02:42.63 UT, the GBM
triggered on and localized GRB~090323 (trigger 259459364/090323002). 
Significant emission was observed in all NaI detectors
on the same spacecraft side (NaI 6 to NaI 11) up to
500 keV. Moreover, BGO detector 1 detected the burst
up to about 3 MeV. Due to the repointing maneuver, the detectors'
angles with respect to the burst location had to be carefully
examined over the whole burst duration. For some detectors, 
the angle changed over about 13$^\circ$. This is a crucial
step before computing the correct detector response
matrices (DRMs) for each individual detector
over a particular burst interval.
The GRB light curve is shown in Figure \ref{Fig_LC_090323} ({\it left panel})
and is characterized by a first group of 
peaks showing a lot of substructure
between the trigger time (T$_0$) and T$_0$+70~s, a plateau with very little emission
from T$_0$+70~s to $\sim$T$_0$+110~s and finally three late well--defined
and equally--spaced peaks, each about 10~s long. 
By combining the orientation changes with the 
peculiarities of the GRB profile,
the burst was divided in seven intervals ({\bf a} to {\bf g}) for the spectral analysis.
The burst characteristic durations T$_{90}$
and T$_{50}$, which represent the time during 
which 90\% and 50\% of the event flux
is collected \cite{KOU93}, were found to be 
133.1$\pm$1.4~s and 42$\pm$4~s, respectively,
where the error bars reflect the 1--sigma
statistical uncertainties.

On 2009 March 28 at 09:36:46.51 UT, the GBM
triggered on and localized GRB~090328
(trigger 259925808/090328401). 
Significant emission
was observed in nine NaI detectors, 
including all detectors on one
side of the spacecraft (NaI 6 to NaI 11) plus NaI 3 and NaI 4.
The burst was detected by BGO detector 1 up to about 2 MeV.
NaI 7 and NaI 8 are relatively stable in orientation during
the spacecraft slew (the angles change over $\sim$4$^\circ$ only) and represent the 
best choice for the spectral analysis.
Light curves for GRB~090328 are shown in the {\it right panel}
of Figure \ref{Fig_LC_090323}.
The burst profile is similar to the one of GRB~090323,
displaying three emission peaks followed by a decreasing 
plateau phase and final peak which is mainly
detected in the NaI detectors.
However, GRB~090328 has a much shorter
duration with T$_{90}$ = 57$\pm$3~s and T$_{50}$ = 15.4$\pm$1.0~s.
For the spectral analysis, the burst was divided 
in five intervals ({\bf a} to {\bf e}).
\section{Time--resolved spectral analysis}
Spectral analysis was performed for both GRBs using the GBM data only.
The NaI data are usually fit from 8 keV to 1 MeV and the BGO data
from 250 keV to 40 MeV using the TTE data type \cite{MEE09}.
The fits were performed with the spectral analysis
software package RMFIT (version 3.2).
For the spectral analysis, the best NaI
detector pair was fitted together with detector BGO 1
with various spectral models, i.e. (i) a simple power--law (PL) function;
(ii) a power--law function with 
an exponential high--energy cutoff (Comptonized), where
the cutoff energy is parameterized as E$_{\rm peak}$; and (iii)
a typical GRB Band function \cite{BAN93}.
The time--integrated spectrum of GRB~090323 is best modeled by a Band
function with an $E_{\rm peak}$ of about 600 keV. 
In the case of GRB~090328, the time--integrated spectrum
is best modeled by a power--law with exponential
high-energy cutoff with an $E_{\rm peak}$ of about 750 keV. 

\renewcommand{\tabcolsep}{2pt}
\begin{figure}[t!]
\centering
\begin{tabular}{cc}
\includegraphics[width=0.48\textwidth,bb=-30 0 540 690,clip]{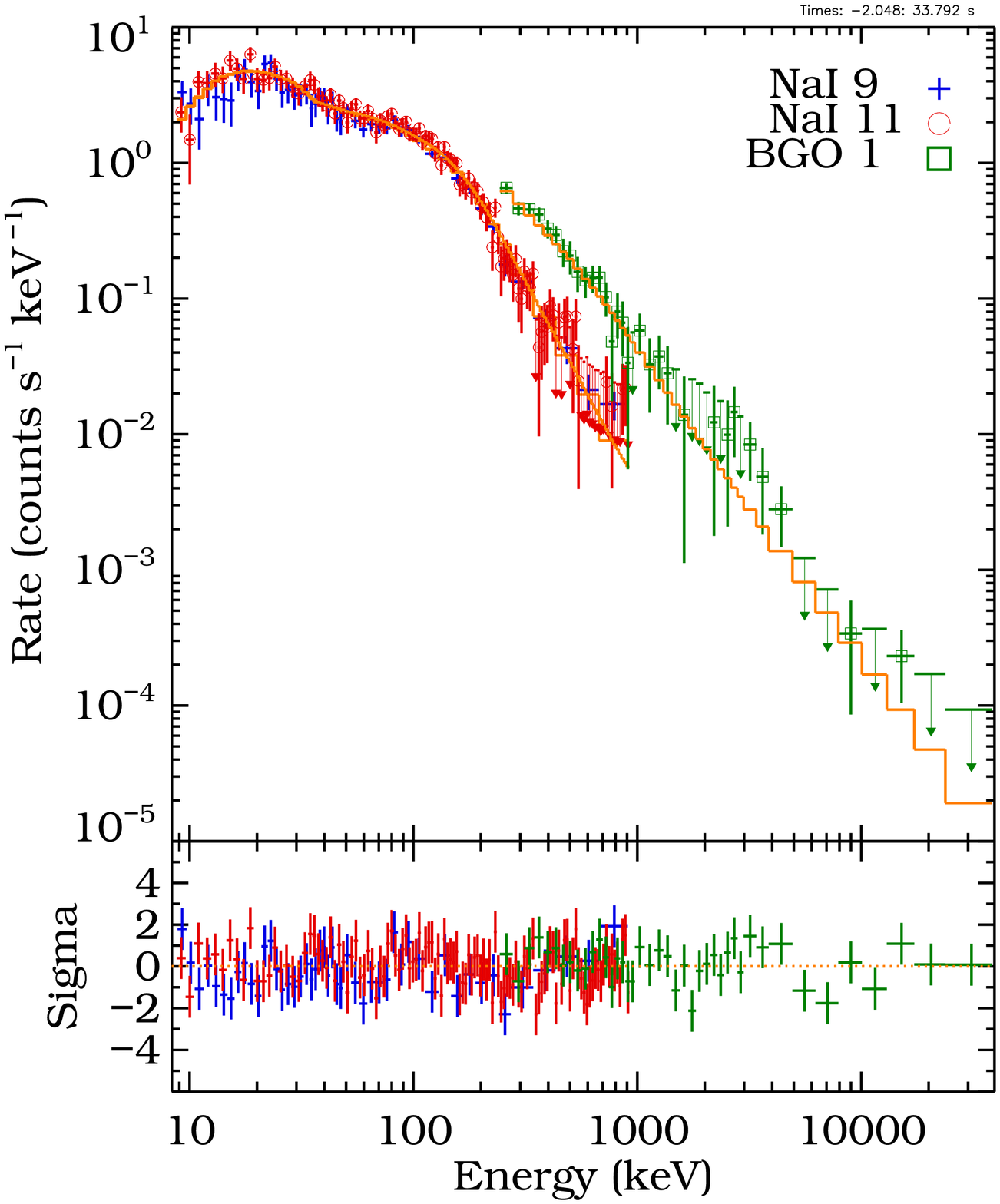} & 
\includegraphics[width=0.48\textwidth,bb=-30 0 540 690,clip]{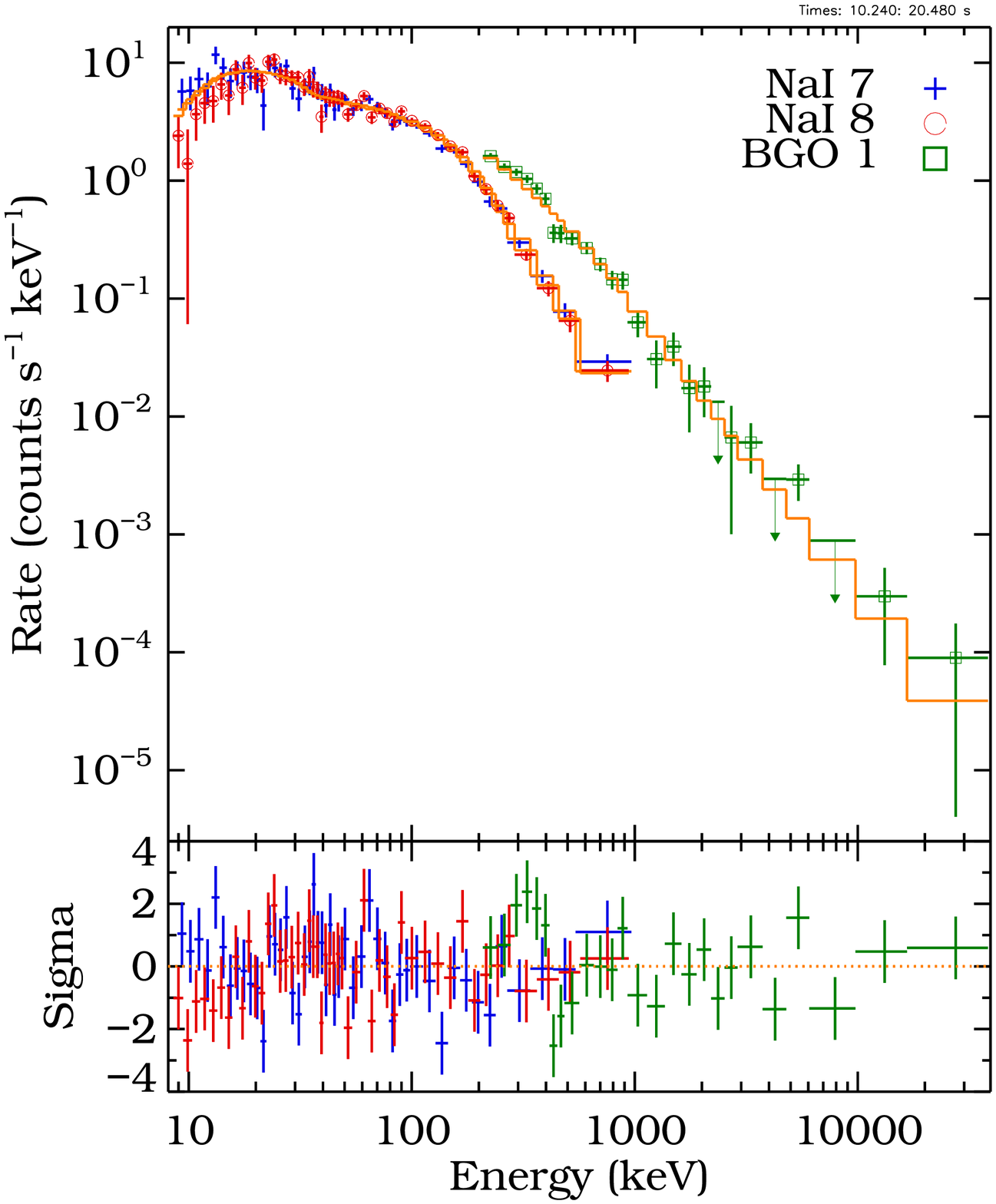}
\end{tabular}
\caption{
{\it Left:} Band fit of GBM data to interval {\bf a} of GRB~090323 
(from T$_0$-2.0~s to T$_0$+33.7~s);
{\it Right:} Band fit of GBM data to interval {\bf b} of GRB~090328 
(from T$_0$+10.2~s to T$_0$+20.5~s).
}
\label{Fig_Spec}
\end{figure}

Results for the time--resolved spectral analysis of 
GRB~090323 and GRB~090328 are given in Table \ref{Tab_Spec}.
Spectral evolution throughout both GRBs 
is apparent from the changing $E_{\rm peak}$
values. The highest $E_{\rm peak}$ is measured during the 
first emission episode (interval {\bf a}) for GRB~090323
and during the second emission episode (interval {\bf b})
for GRB~090328. Both intervals are 
best fitted by a Band model. The counts spectra are also
shown in Figure \ref{Fig_Spec}.

\renewcommand{\tabcolsep}{4pt}
\renewcommand{\baselinestretch}{0.7}
\begin{deluxetable}{cccccccccc}
\tabletypesize{\scriptsize}
\rotate
\tablecaption{Fit parameters for the time-resolved spectral fits of GRB~090323 and GRB~090328. \label{Tab_Spec}}
\tablewidth{0pt}
\tablehead{
\colhead{Interval} & 
\colhead{Det.}     & 
\colhead{Time range}    & 
\colhead{Model}    & 
\colhead{E$_{\rm peak}^{\rm a}$} & 
\colhead{$\alpha^{\rm b}$} &
\colhead{$\beta^{\rm c}$}  & 
\colhead{$\gamma^{\rm d}$}  & 
\colhead{CSTAT/} & 
\colhead{Energy fluence}\\
 &  & \colhead{(s)} &  & \colhead{(keV)} & & & & \colhead{DOF} & \colhead{(erg\,cm$^{-2}$, 8\,keV--1\,MeV)}
}
\startdata
\multicolumn{4}{l}{\bf {\small GRB~090323}} & & & & & & \\
\hline\noalign{\smallskip}
$\cdots$& n9,b1    &  $\;\;\;$-2.0--150.5 & {\it Band} & 591 ($^{+36}_{-33}$) & -1.05 ($\pm$0.02) & -2.7 ($^{+0.2}_{-0.4}$) & $\cdots$ & 888/237 & (1.22 $\pm$ 0.02)$\times10^{-4}$ \\  
{\bf a.}& n9,nb,b1 &  $\,$-2.0--33.8  & {\it Band} &  710 ($^{+78}_{-67}$) & -0.94 ($\pm$0.03) & -2.3 ($^{+0.1}_{-0.2}$) & $\cdots$ & 474/356 & (3.48 $\pm$ 0.06)$\times10^{-5}$ \\  
{\bf b.}& n9,nb,b1 &  33.8--56.3  & {\it Comp} & 447 ($^{+25}_{-23}$) & -0.81 ($\pm$0.03) &  $\cdots$ & $\cdots$ & 517/357 & (2.72 $\pm$ 0.04)$\times10^{-5}$ \\
{\bf c.}& n9,nb,b1 &  56.3--73.7  & {\it Comp} & 521 ($^{+23}_{-22}$) & -0.84 ($\pm$0.02) &  $\cdots$ & $\cdots$ & 475/357 & (3.47 $\pm$ 0.04)$\times10^{-5}$ \\
{\bf d.}& n9,nb,b1 & $\;\;$73.7--112.6 & {\it PL} &  $\cdots$         & $\cdots$ & $\cdots$  & -1.57 ($\pm$0.02) & 663/359 & (1.06 $\pm$ 0.02)$\times10^{-5}$ \\
{\bf e.}& n7,n9,b1 & 112.6--127.0 & {\it Comp} & 200 ($^{+40}_{-30}$) & -1.18 ($\pm$0.09) &  $\cdots$ & $\cdots$ & 437/358 & (3.76 $\pm$ 0.22)$\times10^{-6}$ \\
{\bf f.}& n7,n9,b1 & 127.0--136.2 & {\it Comp} & 154 ($^{+34}_{-22}$) & -1.07 ($^{+0.14}_{-0.13}$) & $\cdots$ & $\cdots$ & 371/358  & (2.34 $\pm$ 0.14)$\times10^{-7}$ \\
{\bf g.}& n6,n9,b1 & 136.2--150.5 & {\it Band} & 108 ($^{+13}_{-14}$) & -1.00 ($^{+0.11}_{-0.09}$) & -2.3 ($\pm$0.2) & $\cdots$ & 450/358  &  (5.99 $\pm$ 0.22)$\times10^{-6}$  \\
\noalign{\smallskip}\hline\noalign{\smallskip}
\multicolumn{4}{l}{\bf {\small GRB~090328}} & & & & & & \\
\hline\noalign{\smallskip}
$\cdots$& n7,n8,b1    &  $\;\;\;$0.0--66.6 &  {\it Comp} &  744 ($^{+50}_{-47}$) & -1.07 ($\pm$0.02) & $\cdots$ & $\cdots$ & 644/360 & (5.09 $\pm$ 0.04)$\times10^{-5}$ \\  
{\bf a.} & n7,n8,b1 &  0.0--10.2 & {\it Comp} & 703 ($^{+71}_{-61}$) & -0.80 ($\pm$0.04) & $\cdots$ & $\cdots$ & 388/361 & (1.20 $\pm$ 0.02)$\times10^{-5}$ \\
{\bf b.} & n7,n8,b1 & 10.2--20.5 & {\it Band} & 716 ($^{+56}_{-52}$) & -0.84 ($\pm$0.03) & -2.4 ($^{+0.1}_{-0.2}$) & $\cdots$ & 430/360 & (2.01 $\pm$ 0.03)$\times10^{-5}$\\
{\bf c.} & n7,n8,b1 & 20.5--30.7 & {\it Band} & 451 ($\pm$66)        & -1.07 ($^{+0.05}_{-0.04}$) & -2.2 ($^{+0.2}_{-0.3}$)  & $\cdots$ & 436/360 & (1.22 $\pm$ 0.02)$\times10^{-5}$\\
{\bf d.} & n7,n8,b1 & 30.7--55.3 & {\it PL}   & $\cdots$             &   $\cdots$ & $\cdots$ & -1.52 ($\pm$0.03)          & 480/362 & (4.12 $\pm$ 0.15)$\times10^{-6}$ \\ 
{\bf e.} & n7,n8,b1 & 55.3--66.6 & {\it Comp} & 123 ($^{+33}_{-21}$) & -1.41 ($^{+0.12}_{-0.11}$) & $\cdots$ & $\cdots$ & 354/361 & (1.87 $\pm$ 0.12)$\times10^{-6}$
\enddata
\tablecomments{The time range values are relative to the trigger time $T_0$. \\
$^{\rm a}$ Fitted $E_{\rm peak}$ for the Band or Comptonized models. \\
$^{\rm b}$ Low--energy spectral index $\alpha$ for the Band or Comptonized models. \\
$^{\rm c}$ High--energy spectral index $\beta$ for the Band model. \\
$^{\rm d}$ Spectral index $\gamma$ for the PL model.
}
\end{deluxetable}



\bibliographystyle{aipprocl} 

\begin{thebibliography}{}
  \bibitem{ATW09} ATWOOD W.~B.        ET AL.,   {\it ApJ}          {\bf 697}  (2009) 1071.
	\bibitem{MEE09} MEEGAN C.~A.        ET AL.,   {\it ApJ}          {\bf 702}  (2009) 791.
  \bibitem{CHO09} CHORNOCK R.         ET AL.,   {\it GCN Circular} {\bf 9028} (2009).
  \bibitem{CEN09} CENKO S.~B.         ET AL.,   {\it GCN Circular} {\bf 9053} (2009).
  \bibitem{ABD10} ABDO A.~A.          ET AL.,   {\it ApJ in preparation}      (2010).
	\bibitem{KOU93} KOUVELIOTOU C.      ET AL.,   {\it ApJ}          {\bf 413}  (1993) L101.
	\bibitem{BAN93} BAND D.             ET AL.,   {\it ApJ}          {\bf 413}  (1993) 281.
\end{thebibliography}

\end{document}